\documentclass[nohyper,12pt,letterpaper]{JHEP}
\usepackage{epsfig}
\newcommand\eqref[1]{(\ref{#1})}
\newfont{\frak}{eufm10 scaled 1200}

\newfont{\Bbb}{msbm10 scaled 1200}     %instead of eusb10
\newcommand{\mathbb}[1]{\mbox{\Bbb #1}}
\DeclareSymbolFont{AMSa}{U}{msa}{m}{n}
\DeclareSymbolFont{AMSb}{U}{msb}{m}{n}
\let\Box\relax
\DeclareMathSymbol{\Box}{\mathord}{AMSa}{"03}

\title{The Acceleration of the Universe, a Challenge for String Theory }

\author{W. Fischler, A. Kashani-Poor, R. McNees, S.
Paban, \\
  Department of Physics\\
  University of Texas, Austin, TX 78712\\
E-mail: \email{fischler, kashani, mcnees \\
\hspace{1.2cm}paban, @physics.utexas.edu}}

\abstract{Recent astronomical observations indicate that the universe is
accelerating. We argue that generic quintessence models that accomodate
the present day acceleration tend to accelerate eternally. As a
consequence the resulting spacetimes exhibit event horizons. Hence,
quintessence poses the same problems for string theory as asymptotic de
Sitter spaces.}


\preprint{\hepth{0104181}\\ UTTG-07-01 \\}
\begin{document}

%%%%%%%%%%%%%%%%%%%%%%%%%%%%%%%%%%%%%%%%%%%%%%%%%%%%%%%%%%%%%%%%%%%%%%%%%%%%
%          Table of contents automatic !!!                                 %
%%%%%%%%%%%%%%%%%%%%%%%%%%%%%%%%%%%%%%%%%%%%%%%%%%%%%%%%%%%%%%%%%%%%%%%%%%%%

\section{Introduction}

Recent observations of supernovae indicate that the universe is accelerating \cite{Perlmutter:1997ds}
\cite{Riess:1998cb}. This suggests that, at present, the energy density of the universe is dominated by
a fluid with equation of state $p < - \frac{1}{3} \; \rho$. If this inequality persists for all future
times, the universe will eternally accelerate and hence exhibit an event horizon. The holographic
principle then implies that a finite dimensional Hilbert space is sufficient to describe the universe
\cite{Fischler:1998st} \cite{Bousso:2000nf} \cite{Bousso:2000md} \cite{Bousso:1999xy}. This peculiar
property has recently been discussed in the literature, in the context of asymptotic de Sitter
cosmologies, by Banks \cite{Banks:2000fe} and elaborated on by others \cite{Banks:2001yp}. The same
line of argument presented by these authors can be repeated in the more general context of an eternally
accelerating universe.

A universe with an event horizon poses serious challenges for string (M-)
theory (we will use the name `string theory' in the rest of the paper for the Underlying Theory). 
 It is quite a difficult task to find a string theory
that manages somehow to have a finite number of states. Achieving this
goal would still leave the formidable task of formulating meaningful
questions, i.e. finding observables, within string theory in a universe with an event
horizon. Constructing a conventional S-matrix is not
possible because the local observer inside his horizon is not able to
isolate particles to be scattered. One is left with the problem of
finding alternatives to a conventional S-matrix; for recent proposals, see~\cite{Banks:2001yp}~\cite{Witten:strings2001}.

Let us return to the claim that, at present,
the universe is accelerating. For the sake of argument we will
assume that this result will be confirmed.
  This alone does not imply that the universe will
accelerate indefinitely.
In the case of an asymptotic de Sitter universe indefinite acceleration does indeed take place, leading to the
problems mentioned above. Quintessence \cite{Ratra:1988rm} \cite{Wang:2000fa} offers a possible explanation for the observed acceleration of the universe without invoking a cosmological constant.
From a string theory point of view, quintessence at first sight seems more palatable than having a true cosmological constant, as
scenarios in string theory can be entertained where moduli
 play the role of quintessence. 

The claim we make in this short paper is
that, quite generically, quintessence also leads to cosmologies with event horizons and hence carries with it the same challenges
for string theory as does a cosmological constant. We discuss some possible loopholes to this conclusion.

 The paper is organized as follows: in section 2, we discuss the
 general properties of an accelerating universe. In section 3, we
consider several realizations of quintessence. 
We conclude in the last section by comparing the implications of quintessence for string theory with those of a
cosmological constant.

\section{Brief history of an accelerating universe}

The universe at large scales and in the recent cosmological past can be described by
a flat Friedman-Robertson-Walker universe
$$ ds^2 = - dt^2 + a^2(t)(dr^2 +r^2 d\Omega^2) \,,$$
with a perfect fluid as matter content. Einstein's field equations for this setup yield
\begin{eqnarray}
3{(\dot{a})^2 \over a^2} &=& 8 \pi \rho(a)   \label{frw}  \\
3 {\ddot{a} \over a} &=& -4 \pi (\rho + 3 p) \,.  \label{srw}
\end{eqnarray}
Here, $\rho$ and $p$ are the energy density and the pressure of the perfect fluid that fills the universe (the second equation, incidently, follows from the first and the thermodynamic relation $dE= -pdV$, valid for isentropic expansion).

From equation (\ref{srw}), we can read off the condition for the expansion of the universe to be accelerating:

\begin{equation}
p < -{1\over3}\rho \,. \label{cau}
\end{equation}
An eternally accelerating universe exhibits an event horizon: there exist regions of spacetime that remain inaccessible to the geodesic observer at arbitrary late times. In FRW spaces, this occurs whenever light rays, emitted at time $t_0$ , only travel a finite coordinate distance throughout the remaining history of the universe, i.e. whenever the integral

$$ \Delta= \int_{t_0}^{\infty} {dt\over a(t)} $$
converges: events whose coordinates at time $t_0$ lie beyond $\Delta$ can then never communicate with the observer at $r=0$.

Let us assume an ideal fluid with the equation of state $p = \kappa \rho$, $\kappa$ fixed. It is not hard to solve Einstein's equations in this case to obtain

\begin{equation}
\rho \sim a^{-3(1+\kappa)}  \label{density}
\end{equation}
and thus

\begin{equation}
a(t)\sim t^{2\over3(1+\kappa)} \,.  \label{scale factor}
\end{equation}
We see that the condition for the existence of an event horizon is $\kappa < - {1 \over 3}$, thus coinciding with the condition for an accelerating universe, equation (\ref{cau}).
 \FIGURE{
	 \epsfig{file=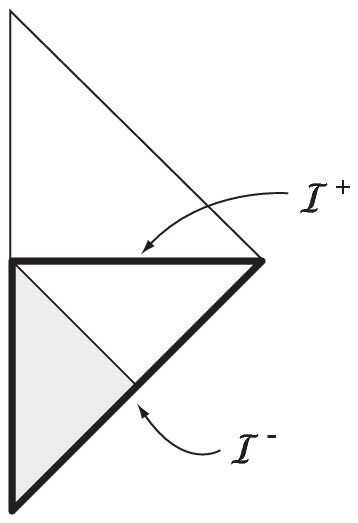}
	 \caption{\label{penrose} The lower triangle is the Penrose diagram for flat FRW space with $\kappa < -{1 \over 3}$. The shaded triangle is the region accessible to the observer at $r=0$.}
        } 
The same argument can be made with the help of Penrose diagrams. FRW spaces are conformally flat, i.e. their metric can be written as $ ds^2 = a(\tau)^2(-d\tau ^2  + dr^2 + r^2 d\Omega^2) $. Their Penrose diagram can therefore be embedded in that of Minkowski space. Conformal time $\tau$ is determined by the differential equation

\begin{equation}
{d \tau \over dt} = {1 \over a(t)} \,.
\end{equation}
Cosmological time $t$ ranges from $0$ to $\infty$. Fixing $\tau(t_0)=\tau_0$ at an arbitrary time $t_0$, $\tau(0)$ and $\tau(\infty)$ are determined by the integrals:

\begin{eqnarray}
\tau(0) = \tau_0 - \int_{0}^{t_0} {1 \over a(t)}   \\  \label{past infinity}
\tau(\infty) = \tau_0 + \int_{t_0}^{\infty} {1 \over a(t)}  \,.  \label{future infinity}
\end{eqnarray}
With the scale factor given by equation (\ref{scale factor}), $\kappa < - {1 \over 3}$ implies the divergence of the first and the convergence of the second integral (and vice versa for $\kappa > -{1\over3}$). Therefore, with an appropriate choice of $\tau_0$, $\tau$ can be made to range, for the case of this eternally accelerating universe, from $-\infty$ to $0$. The Penrose diagram for this FRW space hence covers the lower triangle of the Penrose diagram for Minkowski space, as indicated in figure \ref{penrose}. The existence of an event horizon can now be read off from the fact that $\cal{I}^+$ is spacelike.

The model we have been considering so far does not coincide with the cosmological Standard Model, in which the universe was dominated by radiation and then by matter in the far past, certainly not by a $\kappa < - {1 \over 3}$ fluid. For the purposes of this paper, we are interested in the resulting modifications to our analysis only to the extent in which the conformal structure of spacetime is altered. It therefore suffices to add a second component to our fluid, say with $\kappa = {1 \over 3}$ for radiation (in fact any $\kappa > -{1 \over 3}$ component would have the same effect):

\begin{eqnarray}
p = {1 \over 3} \rho_r + \kappa \rho_q             \,.
\end{eqnarray}

 \FIGURE{
	 \epsfig{file=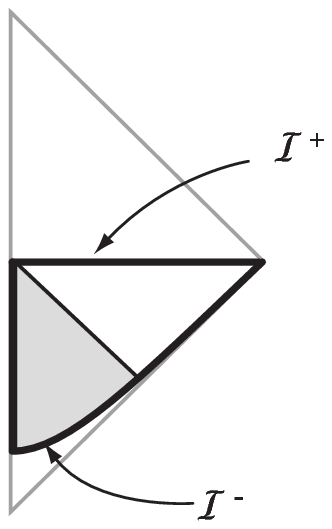}
	 \caption{\label{penrose2} Adding a radiation component renders $\cal{I}^-$ spacelike. The shaded triangle is the region accessible to the observer at $r=0$.}
        } 

Solving the equations (\ref{frw}) and (\ref{srw}) for this case, we see that the overall energy density is

\begin{eqnarray}
\rho = {c_{r} \over a^4} + {c_{q}  \over a^{3(1 + \kappa)}}   \,.
\end{eqnarray}
With the initial condition $a(0) = 0$, the radiation component will dominate at early times, but then redshift away more quickly than the $\kappa < -\frac{1}{3}$ component. The behavior of the scale factor will thus be dictated by $\rho_r$ at early times and by $\rho_q$ at late times:

\begin{eqnarray}
a_{\mathrm{early}}(t) &\sim& t^{1 \over 2} \\
a_{\mathrm{late}}(t) &\sim& t^{2\over3(1+\kappa)} \,.
\end{eqnarray}
Repeating our previous analysis, we find that the total duration of the universe in conformal time in this spacetime is finite. The corresponding Penrose diagram, embedded in that of Minkowski space, is depicted in figure \ref{penrose2}. Now, both $\cal{I}^-$ and $\cal{I}^+$ are spacelike. The presence of an event horizon is, of course, not altered by this modification.

Does the existence of an event horizon imply the presence of Hawking radiation? The entire accelerated universe is conformally equivalent to the steady-state universe (i.e. spatially flat de Sitter space).
In the case of a purely $\kappa < - {1 \over 3}$ fluid (figure \ref{penrose}), the region of the Penrose diagram accessible to a local observer at $r=0$ is conformally equivalent to static de Sitter space. The discussion in \cite{Candelas:1979gf} hence suggests that the analysis for de Sitter space should equally well go through for our accelerated FRW spaces. The more realistic model (figure \ref{penrose2}) is conformally equivalent to an asymptotically de Sitter Big Bang model. Therefore, by the same token, the existence of a thermal heat bath in asymptotic de Sitter space would imply the existence of a thermal heat bath in this case as well. The existence of a temperature would be another impediment to constructing an S-matrix. A local observer does not see free particles near the horizon, rather they appear immersed in a bath of photons which are highly energetic as a result of blue-shifting.

\section{Quintessence}

	Quintessence endeavors to replace an inert cosmological constant
with a dynamic negative pressure component. For the purpose of this paper,
quintessence will refer to a minimally coupled (pseudo)scalar field with a
potential that decreases as the field increases. The field couples to gravity
through a Lagrangian of the form:
  \begin{eqnarray}
	\mathcal{L} & = & \sqrt{g} \; \left(\,\frac{1}{2}\; (\partial \phi)^{2}
	 + V(\phi) \right) \label{eq:lagrangian}  \,.
  \end{eqnarray} 
The stress-energy tensor can be cast in the perfect fluid form. If we assume that
the field is spatially homogenous then the energy density and pressure are given
by:
 \begin{eqnarray}
	\rho & = & \frac{1}{2}\; \dot \phi^{\,2} + V(\phi) \\
	p & = & \frac{1}{2}\; \dot \phi^{\,2} - V(\phi) \,.
\end{eqnarray}
The equation of state then takes the form:
\begin{eqnarray}
	\kappa & = & \frac{\dot \phi^{\,2} - 2 \, V(\phi)}{\dot \phi^{\,2} + 2
	\, V(\phi) } \,.
\end{eqnarray}
The equation of state thus ranges over $-1 < \kappa < 1$, depending on the
dynamics of the field. When the kinetic energy dominates the potential, the
equation of state is such that $\kappa \sim 1$. In this case, the arguments in
the last section tell us that the energy density is red-shifted away at the
maximum rate consistent with causality, $\rho_{\phi} \sim a^{-6} $.
When $\dot \phi^{\,2} < 2 \, V(\phi)$, the factor $\kappa$ falls below $0$ and the
field acts as a negative pressure component. If $\dot \phi^{\,2} < V(\phi)$ then
$\kappa<-\frac{1}{3}$ and the field contributes to the acceleration of the
universe. Quintessence models, to be consistent with current observation, must evolve
into this last regime where the dominance of the potential energy leads to
acceleration. Our point is that the most commonly studied models tend to remain in this regime eternally, thus yielding indefinitely accelerating spacetimes with all the complications discussed above. To demonstrate this last point, we present analytical and numerical
analyses of three common quintessence models that lead to a perpetually
accelerating universe.

	To study the evolution of quintessence scenarios, we solve, either
analytically or numerically, the coupled Einstein equations for the field $\phi$
and the scale factor $a$. For a Robertson-Walker spacetime the
Lagrangian~\eqref{eq:lagrangian} gives the equation of motion:
  \begin{eqnarray}
	\ddot \phi + 3 \, H \, \dot \phi + V'(\phi) & = & 0  \,.  \label{eq:eom}
  \end{eqnarray}
This equation, combined with the Hubble equation (\ref{frw}), determines the evolution
of the universe.

We begin by reviewing a discussion presented by Ratra and Peebles \cite{Ratra:1988rm}, in which the potential for the scalar field consistent with the equation of state $p = \kappa \rho$ for fixed $\kappa$ is derived. This condition allows us to express $\dot{\phi}^2$ and $V(\phi)$ in terms of $\rho$:
\begin{eqnarray}
\dot{\phi}^2 &=& (1+\kappa) \rho     \label{phidot} \\
V(\phi) &=& {1 \over 2} (1- \kappa) \rho \,.
\end{eqnarray}
Substituting expression (\ref{phidot}), solved for $\rho$, into Einstein's equation (\ref{frw}) gives a differential equation for $\phi(a)$ \cite{Ratra:1988rm}
\begin{equation}
{d \phi \over da}= { \sqrt{6(\kappa +1)} \over a} \,,
\end{equation}
which can be trivially solved. Incorporating equation (\ref{density}), which relates the density to the scale factor, we find the potential

\begin{eqnarray}
V(\phi) & \sim & \exp \left( -\sqrt{\frac{3}{2}\;(\kappa+1)} \, \phi \right)  \,.
\end{eqnarray}
Given this solution, Ratra et Peebles demonstrate that a special solution for the scalar field and the scale factor indeed exists such that the equation of state $p = \kappa \rho$ is satisfied. Using linear analysis, they demonstrate that for $\kappa < -\frac{1}{3}$, nearby solutions asymptote to this special solution for large $t$. If we choose $\kappa < -\frac{1}{3}$ to guarantee acceleration at our present era, this class of potentials thus generically leads to eternal acceleration.

\begin{figure}
	\begin{center}
   	  \epsfig{file=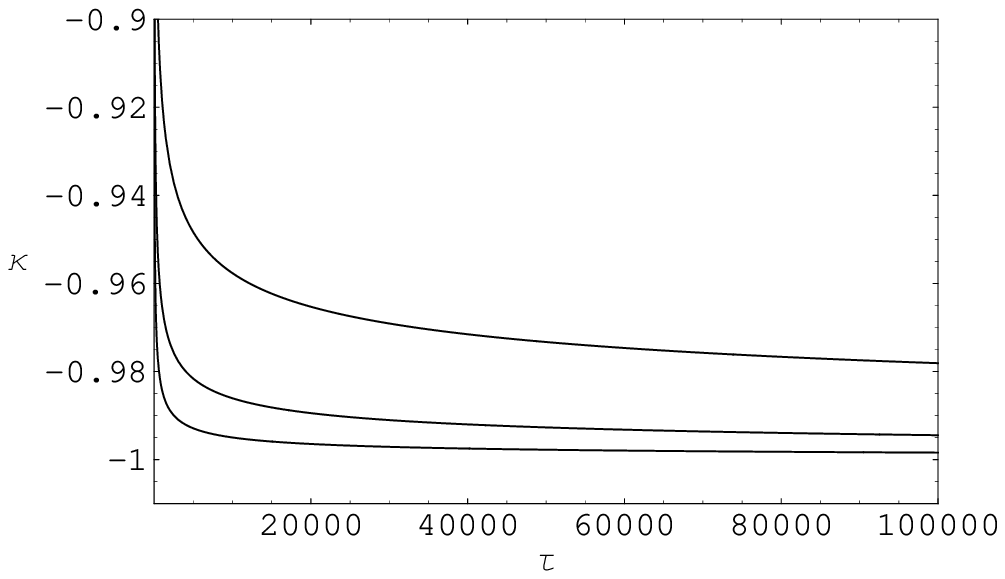}
 	  \caption{\label{fig:inverse}
		   Asymptotic behavior of the equation of state for various
		   inverse power law potentials. The equation of state is taken
		   to be $\kappa \sim -0.6$ at present. Measurements of 
		   $\Omega_{m}$ and properties of tracking solutions fix the
 	           initial conditions. $\tau$ is measured in units of
		   1/$H_0$.}
	\end{center}
  \end{figure}

A second scenario which leads to a perpetually accelerating universe
involves inverse power law potentials of the form:
  \begin{eqnarray}
	V(\phi) & = & \frac{M^{4+n}}{\phi^{n}}  \,.
  \end{eqnarray}
These models were originally studied in \cite{Ratra:1988rm}. Later, they were extensively
studied in \cite{Steinhardt:1999nw}, where the authors demonstrated the existence of `tracker
solutions' which approach a common cosmic evolutionary path from a wide range of
initial conditions. Using the analytical techniques of \cite{Ratra:1988rm} or solving
equations~\eqref{eq:eom} and (\ref{frw}) numerically demonstrates that these potentials
result in solutions which asymptotically approach an equation of state $p=-\rho$,
guaranteeing an eternally accelerating universe. Figure~\ref{fig:inverse} shows
the distant future behavior of the equation of state for various inverse power
law potentials.

\begin{figure}                                                           
        \begin{center}
          \epsfig{file=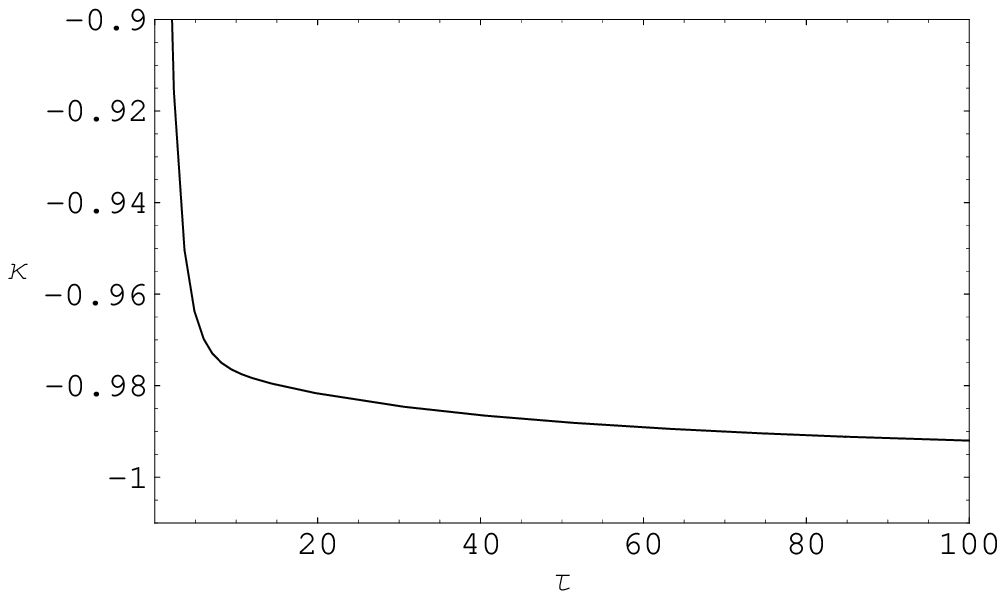} 
          \caption{\label{fig:exppot} Asymptotic behavior of the equation of state for the
	   	   potential~\eqref{eq:exppot}. $\tau$ is measured in units of
                   1/$H_0$.}
        \end{center}
  \end{figure}

	Many of the arguments that apply to inverse power law potentials can be
extended to potentials which are functions of inverse powers
of $\phi$. An example is the exponential potential studied in \cite{Zlatev:1999tr}:
  \begin{eqnarray}
	V(\phi) & = & M^{4} \, \left(\exp(\frac{M_{P}}{\phi})-1  
	\right) \,.   \label{eq:exppot}
  \end{eqnarray}
A direct numerical analysis of the equations for $\phi$ and $a$ shows that
solutions of these models also asymptote towards $p = -\rho$, again
resulting in an accelerating universe. One example is plotted in
Figure~\ref{fig:exppot}.
 
The last two potentials belong to a wider class of potentials referred to as
`runaway scalar fields' \cite{Steinhardt:web}. For a runaway scalar field, $V$, $V'$, $V''$ all approach $0$ as $\phi \rightarrow
\infty$. Also, the ratios $V' / V \rightarrow 0$
and $V'' / V \rightarrow 0$ as well. In addition to the analytic and numerical methods mentioned
above, there is an argument by Steinhardt \cite{Steinhardt:web} that states that a runaway scalar
field insures the eventual acceleration of the universe. Steinhardt argues that
there exists a value of $\phi$, the `sticking point', beyond which the Hubble
expansion overdamps the kinetic energy of the field, leading to a negative
pressure component which eventually dominates matter and radiation and leads to
an accelerating universe.

\vspace{0.4cm}

There are various loopholes that provide an alternative to a perpetually
accelerating universe. One possibility is that the potential does not
decrease forever, but reaches a minimum at some finite $\phi_{0}$ where
$V(\phi_{0})=0$ and $V''(\phi_{0}) > 0$. The field's current slow roll
towards this minimum emulates a cosmological constant. However, once the
field reaches $\phi_{0}$ it undergoes damped oscillations. This behavior
corresponds to the equation of state for non-relativistic matter, $p=0$,
leading to a decelerating universe.

Another possibility is that the potential $V(\phi)$ decreases to a minimum 
$V(\phi_{0})=0$ and then becomes flat for $\phi > \phi_{0}$. This is equivalent
to domination by kinetic energy, resulting in a decelerating universe with
equation of state $p=\rho$.

Yet another possibility is that the scalar field eventually decays into two
photons. A coupling of the form $\phi \, F_{\mu\nu} \, F^{\mu\nu}$ would induce a time dependence of the fine
structure constant. Current limits set by studies of the `Oklo Natural
Reactor' suggest a bound  (\cite{Damour:1996xx} and references therein) as tight as:
  \begin{eqnarray}
	| \dot \alpha / \alpha | & < & 10^{-15} \, \mathrm{yr}^{-1}   \,,
  \end{eqnarray} 
implying that the $\phi \, F_{\mu\nu} \, F^{\mu\nu}$
coupling would be unnaturally suppressed. We might try an axion-like
coupling instead, $\phi \, F_{\mu\nu} \, \tilde F^{\mu\nu}$. This
interaction is constrained by astrophysical considerations, albeit not as
stringently as the previous one. However, if we choose to conserve parity,
the axial-coupling implies that $\phi$ is a pseudo scalar, so we must now
also worry about long-range spin-spin forces.  In a nutshell, such a
decay, while possibly not ruled out, does not present itself naturally.  
For further discussion see~\cite{Witten:2000zk}~\cite{Carroll:1998zi} and
references therein. None of these loopholes seem particularly convincing
as they all require rather special properties from the vantage point of
string theory.

\section{Conclusions}

 The various examples discussed in the previous chapter show that rather
generically, quintessence models result in spacetimes with event horizons,
much like asymptotic de Sitter spaces. A universe endowed with an event
horizon raises challenging questions for string theory. These questions
have been recently elaborated on in the context of asymptotic de Sitter
space~\cite{Banks:2000fe}~\cite{Banks:2001yp}~\cite{Witten:strings2001} .
One can repeat this exercise for a universe that undergoes eternal
acceleration. The main puzzle is what replaces the S-matrix in a universe
with an event horizon, as one does not have the leisure in such a case to
isolate probes from one another or from the thermal bath in which they are
immersed. A related question is what the observables are in a string
theory that is described by a finite dimensional Hilbert space.  We refer
the reader to the previous references for discussions on this frustrating
question.

The observation that the expansion of the universe is
accelerating implies that the dominant form of energy
density today satisfies $p<-{1\over 3}\rho$. Our main message is that
generic quintessence models that satisfy this inequality at present tend to satisfy it eternally. The resulting spacetimes thus exhibit event horizons. Quintessence, very much like a cosmological constant, presents a serious challenge for string theory.

\hspace{1pt}
\hspace{1pt}

\noindent \begin{bf}Note Added:\end{bf} As we were completing this paper,
S. Hellerman, N.~Kaloper and L.~Susskind informed us of their work
\cite{Susskind:2001xx}, in which they reach similar conclusions.

\acknowledgments

This research was supported in part by NSF Grants PHY-0071512 and
PHY-9511632.


\newpage
\begin{thebibliography}{19}

%\cite{Perlmutter:1997ds}
\bibitem{Perlmutter:1997ds}
S.~Perlmutter {\it et al.}  [Supernova Cosmology Project Collaboration],
``Measurements of the Cosmological Parameters $\Omega$ and $\Lambda$ from
the First 7 Supernovae at z $\geq$ 0.35,''
Astrophys.\ J.\ {\bf 483}, 565 (1997)
astro-ph/9608192.
%%CITATION = ASTRO-PH 9608192;%%

%\cite{Riess:1998cb}
\bibitem{Riess:1998cb}
A.~G.~Riess {\it et al.}  [Supernova Search Team Collaboration],
``Observational Evidence from Supernovae for an Accelerating Universe and
a Cosmological Constant,''
Astron.\ J.\  {\bf 116}, 1009 (1998)
astro-ph/9805201.
%%CITATION = ASTRO-PH 9805201;%%

%\cite{Fischler:1998st}
\bibitem{Fischler:1998st}
W.~Fischler and L.~Susskind,
``Holography and cosmology,''
hep-th/9806039.
%%CITATION = HEP-TH 9806039;%%

\bibitem{Bousso:2000nf}
R.~Bousso,
``Positive vacuum energy and the N-bound,''
JHEP{\bf 0011}, 038 (2000)
hep-th/0010252.
%%CITATION = HEP-TH 0010252;%%

%\cite{Bousso:2000md}
\bibitem{Bousso:2000md}
R.~Bousso,
``Bekenstein bounds in de Sitter and flat space,''
hep-th/0012052.
%%CITATION = HEP-TH 0012052;%%
%\cite{Bousso:2000nf}

%\cite{Bousso:1999xy}
\bibitem{Bousso:1999xy}
R.~Bousso,
``A Covariant Entropy Conjecture,''
JHEP{\bf 9907}, 004 (1999)
hep-th/9905177.
%%CITATION = HEP-TH 9905177;%%

%\cite{Banks:2000fe}
\bibitem{Banks:2000fe}
T.~Banks,
``Cosmological breaking of supersymmetry or little Lambda goes back to the future. II,''
hep-th/0007146.
%%CITATION = HEP-TH 0007146;%%

%\cite{Banks:2001yp}
\bibitem{Banks:2001yp}
T.~Banks and W.~Fischler,
``M-theory observables for cosmological space-times,''
hep-th/0102077.
%%CITATION = HEP-TH 0102077;%%

%\cite{Witten:strings2001}
\bibitem{Witten:strings2001}
E.~Witten,
"Quantum Gravity in DeSitter Space,"
Talk at the Strings 2001 Conference,
Tata Institute, Mumbai, India, January 2001.
http://theory.tifr.res.in/strings/

%\cite{Ratra:1988rm}
\bibitem{Ratra:1988rm}
B.~Ratra and P.~J.~Peebles,
``Cosmological Consequences Of A Rolling Homogeneous Scalar Field,''
Phys.\ Rev.\ D {\bf 37}, 3406 (1988).
%%CITATION = PHRVA,D37,3406;%%

%\cite{Wang:2000fa}
\bibitem{Wang:2000fa}
L.~Wang, R.~R.~Caldwell, J.~P.~Ostriker and P.~J.~Steinhardt,
``Cosmic Concordance and Quintessence,''
Astrophys.\ J.\ {\bf 530}, 17 (2000)
astro-ph/9901388.
%%CITATION = ASTRO-PH 9901388;%%

%\cite{Candelas:1979gf}
\bibitem{Candelas:1979gf}
P.~Candelas and J.~S.~Dowker,
``Field Theories On Conformally Related Space-Times: Some Global Considerations,''
Phys.\ Rev.\ D {\bf 19}, 2902 (1979).
%%CITATION = PHRVA,D19,2902;%%

%\cite{Steinhardt:1999nw}
\bibitem{Steinhardt:1999nw}
P.~J.~Steinhardt, L.~Wang and I.~Zlatev,
``Cosmological tracking solutions,''
Phys.\ Rev.\ D {\bf 59}, 123504 (1999)
astro-ph/9812313.
%%CITATION = ASTRO-PH 9812313;%%

%\cite{Zlatev:1999tr}
\bibitem{Zlatev:1999tr}
I.~Zlatev, L.~Wang and P.~J.~Steinhardt,
``Quintessence, Cosmic Coincidence, and the Cosmological Constant,''
Phys.\ Rev.\ Lett.\ {\bf 82}, 896 (1999)
astro-ph/9807002.
%%CITATION = ASTRO-PH 9807002;%%

%\cite{Steinhardt:web}
\bibitem{Steinhardt:web}
P.~J.~Steinhardt,
``Quintessential Cosomology and Cosmic Acceleration,"
http://feynman.princeton.edu/\textasciitilde steinh/

%\cite{Damour:1996xx}
\bibitem{Damour:1996xx}
T.~Damour,
``Gravitation, experiment and cosmology,''
gr-qc/9606079.
%%CITATION = GR-QC 9606079;%%

%\cite{Witten:2000zk}
\bibitem{Witten:2000zk}
E.~Witten,
``The cosmological constant from the viewpoint of string theory,''
hep-ph/0002297.
%%CITATION = HEP-PH 0002297;%%

%\cite{Carroll:1998zi}
\bibitem{Carroll:1998zi}
S.~M.~Carroll,
``Quintessence and the rest of the world,''
Phys.\ Rev.\ Lett.\ {\bf 81}, 3067 (1998)
astro-ph/9806099.
%%CITATION = ASTRO-PH 9806099;%%

%\cite{Susskind:2001xx}
\bibitem{Susskind:2001xx}
S.~Hellerman, N.~Kaloper and L.~Susskind,
``String Theory And Quintessence,''
hep-th/0104180


%-----Citations are then made by \cite{how we call it} in text ------------
%-----\bibitem without [how it is denoted] is numbered 1,2,3....



\end{thebibliography}
\end{document}